\documentclass[twocolumn,pre,showpacs,showkeys,
               amsmath,amssymb,floatfix,10pt]{revtex4}
\usepackage{graphicx}
\usepackage{dcolumn}
\usepackage{bm}
\usepackage{graphicx}
\usepackage[colorlinks=true]{hyperref}
\hypersetup{urlcolor=blue, citecolor=red}

\usepackage{subfigure}
\usepackage{amsmath}
\usepackage{amssymb}
\usepackage{wrapfig}
\usepackage{color}
\usepackage[english]{babel}

\def\<{\langle}
\def\>{\rangle}

\begin{document}

\title{A simple Monte Carlo model for crowd dynamics} 
\author{Francesco Piazza~\footnote{Present address: Centro Interdipartimentale per le 
Dinamiche Complesse (CSDC), Dipartimento di Fisica, Via G. Sansone 1, 50019 Firenze, Italy.}}
\email{Francesco.Piazza@gmail.com}
\affiliation{Ecole Polytechnique F\'ed\'erale de Lausanne, 
Laboratoire de Biophysique Statistique, ITP-SB,
BSP-720, CH-1015 Lausanne, Switzerland
}

\begin{abstract}

In this paper we introduce a simple Monte Carlo method for simulating 
the dynamics of a crowd. Within our model a collection of hard-disk agents is subjected to a 
series of two-stage steps, implying (i) the displacement of one specific agent followed by (ii) a
rearrangement of the rest of the group through a Monte Carlo dynamics. The rules for the combined steps 
are determined by the specific setting of the granular flow, so that our scheme should be 
easily adapted to describe crowd dynamics issues of many sorts, from stampedes in panic scenarios
to organized flow around obstacles or through bottlenecks.
We validate our scheme by computing the serving times statistics of a group of agents 
crowding to be served around a desk. In the case of a size homogeneous crowd, we recover intuitive 
results prompted by physical sense. However, as a further illustration of our theoretical framework, 
we show that heterogeneous systems display a less obvious behavior, as smaller agents feature shorter 
serving times. Finally, we analyze our results in the light of known properties of non-equilibrium 
hard-disk fluids and discuss general implications of our model.   
\end{abstract}
%


%

\maketitle
\section{Introduction}
Crowd dynamics is the object of a comprehensive body of literature, from 
classic broad treatises covering general topics~\cite{Bon:1952oq,Curtis:1992qe,Locher:2001ai,Miller:2000dp} to 
more specific applications, ranging from psycho-dynamical models~\cite{Singh:2009fy} and 
control issues~\cite{Spieser:2009nx,Lmmer:2008eq} to panic scenarios~\cite{Helbing:2000fu,Gwynne:1999oz}.
Likewise, several in-depth reviews discuss computational models of pedestrian 
dynamics~\cite{Nagatani:2002fb,Schreckenberg:2001et,Helbing:2001dz,Castellano:2009jx}, 
ranging from lattice-gas models~\cite{Nagatani:2002fb} to molecular-dynamics-based  simulation schemes 
such as the behavioral (or social) force model~\cite{Helbing:1991jx,Helbing:2001by} and 
cellular automata methods~\cite{:2000wt}.
\\
\indent It is the purpose of the present paper to introduce a novel model of crowd dynamics, based on a
simple Monte Carlo (MC) scheme of hard disks dynamics. Our aim is two-fold: one the one side, we wish to 
outline the essentials of an original modeling strategy, that has never been considered so far to our
knowledge. The simple philosophy behind our model can be effortlessly extended and adapted to 
many different scenarios, providing a viable and fast alternative to many current schemes for 
exploring real-life contexts of many sorts in crowd dynamics issues. 
At the same time, we also apply our theoretical framework to a general problem, that of finding 
the distribution of serving times of individuals forming a crowd before a desk. Our model is
shown to provide results in line with the physical intuition in the case of identical 
agents, which confirms the soundness of our theoretical approach. However, we find less obvious 
results when the case of poly-disperse crowds are considered, which constitutes a first illustration 
of the usefulness of our model.
It should be noted that, while queuing strategy issues have been analyzed 
in some specific contexts of crowd dynamics, such as parallel multi-lane 
pedestrian traffic~\cite{Bantang:2006fv},
surprisingly it seems that no simple computational study has yet tackled the
problem examined in this paper.  
\\
\indent 
The article is organized as follows. First we describe the model and discuss the main 
results that validate our scheme in the case of a homogeneous crowd.
Subsequently, we introduce a given degree of poly-dispersity so as to examine 
the statistics of serving times and their dependence on
agents' size. Then, before summarizing and discussing our findings, we report an instructive 
structural analysis of the two-dimensional crowd-queuing systems.

\section{The model\label{sec:1}}

In our model a crowd is represented by a collection of hard disks, possibly 
enclosed in a given perimeter with specific boundary conditions. Broadly speaking, 
the evolution of the crowd configuration is considered as a two-step process. 
In the first step (i) one agent is chosen according to a given rule and is displaced 
following a certain prescription (possibly to infinity, that is removed from the game). 
In the second step (ii) the rest of the group is let equilibrate through a Monte Carlo dynamics 
under a pre-assigned displacement rule until a stationary state is reached,
as measured through the convergence of a suitable global order parameter.  
The details of the rules applied in steps (i) and (ii) should 
reflect the nature of the process that one wants to simulate. Here we introduce a simple set 
of rules, that could be valid in, or else easily modified for, many circumstances where
one is interested to simulate a flow of people. For example, this could be the case of escape 
dynamics through doors at specific locations, such as in panic 
stampedes~\cite{Helbing:2000fu,Gwynne:1999oz} or other issues related to the motion 
of groups of pedestrians within specified perimeters or across bottlenecks~\cite{Helbing:2001by}.
\\
\indent We consider an ensemble of $N$ impenetrable circles in the plane 
of radii $r_{i} = 1 +(2z-1)\Delta r$, $i=1,2,\dots,N$, where $z$ may be a constant 
(homogeneous crowd), or else be drawn from a set of real numbers specifying the degree of heterogeneity.
In this paper we either set $z=1/2$ or consider a uniform distribution in the interval $z \in [0,1]$. 
In the specific case chosen to validate and illustrate our model, 
dubbed here {\em crowd-queuing}, it should be imagined that the 
disks model individuals waiting around a {\em counter} located at the origin of a 
Cartesian frame of reference and, {\em e.g.}, selling tickets. 
The $N$-agent crowd is served, one individual at a time,  
following the series of serving times $t_{n}$, $n=1,2,\dots,N$.
At each step, the {\em served} agent is removed from the system, so that the group  
at the $n$-th step comprises $N(n)=N-n$ individuals. 
The idea is to model a semi-circular geometry, whereby individuals disappear as soon as 
they are served as they are admitted beneath the counter (say into the theatre or cinema hall), 
{\em e.g.} through a door or down a staircase next to the counter itself. It is clear that, under the reasonable 
hypothesis that walls flanking the counter are reflecting (except for the
aforementioned exit), it is easier to study the equivalent case of a 
fully circular geometry with agents crowding all around 
the teller and being removed from the play as soon as they are 
served~\footnote{We are here explicitly neglecting back-flow effects, associated with the stream 
of served agents getting away from the counter through the crowd, that may 
be relevant in other contexts.}. 
\\
\indent For the sake of simplicity, we will take $t_{n+1}-t_{n}=\tau$ and 
measure time in units of {\em serving steps}.
It should be noted that this implies a minor loss of generality, as we 
are mainly interested in average quantities, such 
as the average time taken to reach the counter from a given 
distance. More generally, in fact, if $P(\tau)$ indicates  the 
(finite-mean) probability density describing the (uncorrelated) 
intervals between serving times,
the average time corresponding to $m$ serving steps will
still be linear in $m$, namely 
$\langle \Delta t \rangle_{m}= m \, \langle \tau \rangle $.
\\
\indent According to our theoretical framework, we approximate the dynamics of 
the group as a sequence of serving and rearrangement steps.
The basic assumption is that the two processes occur on well separated time scales, 
the rearrangement stage being regarded as instantaneous on the time-span set by $\tau$.
At each serving step, the individual that is located closer to the counter disappears,
and the rest of the  agents are rearranged. Such process is iterated 
until all agents have been eliminated. Typically, we record the serving times of all agents 
initially belonging to a certain number of concentric circular shells.
By doing this, we can measure the average number of serving steps
as a function of the initial distance from the counter, the average being taken 
over all agents from the same shells, as well as over
$K$ equivalent initial configurations of the $N$-agent crowd.
\\
\indent In an ordered, one-dimensional queue the rearrangement rule is a trivial 
one - move all agents simultaneously of one position in the direction of the counter.
In our case, agents have the freedom of moving in two
dimensions, whilst they still share a bias towards moving along the radial direction, as 
reaching the counter is  everybody's ultimate goal. In accordance to such principle,
we construct the rule for the Monte Carlo rearrangement stages as follows. At each 
MC step $j$, a radial move is attempted of the same magnitude $\delta r(j)$
for each agent. To that, a component $\delta u(j)$ is added with probability 
$p$ along a randomly chosen unit vector forming an angle with the centripetal direction
in the interval $[-\pi/2,\pi/2]$. The latter moving rule is meant to express the 
agents' will to find alternative routes leading faster to the counter,
taking advantage of local density fluctuations.  
The magnitudes of the two types of attempted displacements  $\delta r(j)$ and $\delta u(j)$ 
are updated at each step, after the 
entire set of moves is performed, so as to keep the acceptance rate $\sigma(j)$  
(that is, the fraction of effectively displaced agents) around the target value $\sigma=0.5$.
The acceptance rule for each MC move is realized by enforcing hard-core repulsion. 
\\
\indent Following the  elimination of the served agent, the MC rearrangement is halted 
when the crowd has reached a {\em stable} configuration. Obviously, there is no unique 
manner to assess whether the system has attained such a status. In fact, also in view 
of the intrinsic non-equilibrium nature of the process that we wish
to simulate, such condition should be understood in a {\em dynamical} sense. For example, 
this can be measured by following the evolution of a suitable global observable $\mathcal{O}$
and fixing the number of rearrangement moves $M$ such that 
$|\overline{\mathcal{O}}_{M} - \overline{\mathcal{O}}_{M-1}|/\overline{\mathcal{O}}_{M}  < {\rm TOL}$,
where $\overline{\mathcal{O}}_{M}= \sum_{k=1}^M {\mathcal{O}}(k \Delta j)/M$ is the 
$M$-step running average computed by sampling every $\Delta j$ MC moves and TOL is 
the required accuracy.
In the calibration runs, we chose to monitor the convergence of the acceptance rate $\sigma$
and of a global {\em structural} indicator, namely the overall jamming index $\Gamma$, 
adapted from Ref.~\cite{Lubachevsky:1990ad} to the case of size-heterogeneous disks,
\begin{equation}
\label{e:jam}
\Gamma = \frac{1}{N} \sum_{i=1}^N \gamma_{i} = 
\frac{1}{N} \sum_{i=1}^N \left[
                          \frac{\sum_{j\in\mathcal{S}_{3}(i)} [d_{ij}-(r_{i}+r_{j})]}
                          {3r_{i} + \sum_{j\in\mathcal{S}_{3}(i)} r_{j}}
                         \right]
\end{equation} 
where $\mathcal{S}_{3}(i)$ is the ensemble comprising the first three neighbors 
of the $i$-th agent and $d_{ij}$ are the center-to-center separations. 
As a results of different calibration tests, we found that the choice  
$\Delta j = 50$, ${\rm TOL}=10^{-4}$ yielded a satisfactory compromise between accuracy
in the convergence of both indicators and simulation speed,
with $M(n)/N(n) \simeq const$ ($n$ being the serving step).
For the sake of computation celerity, we only monitored the observable $\sigma$
for arresting the rearrangement stages in production runs. 
\\
\indent The initial configurations for the serving process 
are obtained as follows. The disks are first arranged on a square lattice within a circle 
of radius $R=1+\Delta r$ ($R=1$ for $z=1/2$) and their radii rescaled so as to match the required 
value of the initial disk area fraction $\phi=\sum_{i=1}^{N} r_{i}^{2}/R^2$.
An initial cycle of $\mathcal{M}_i$ Monte Carlo steps is then performed with 
$p=0$ (purely radial moves) and boundary conditions such that displacements
are taken modulo $2R$. The configurations obtained after different series of 
$\mathcal{M}_i \simeq \mathcal{O}(10^{4})$ MC steps are used as independent 
initials conditions for the serving process. Of course, arbitrary initial geometries 
can be enforced in the same manner, by retaining only the agents that lie 
within the required perimeter. 
\\
\indent It should be noted that, besides the measure of the size heterogeneity $\Delta r$,
our simulation protocol has only one parameter, namely 
the probability $p$ of non-radial displacements.

%
\section{The serving dynamics\label{sec:2}}
%

\begin{figure}[t!]
\begin{center}
\includegraphics[width=\columnwidth,clip]{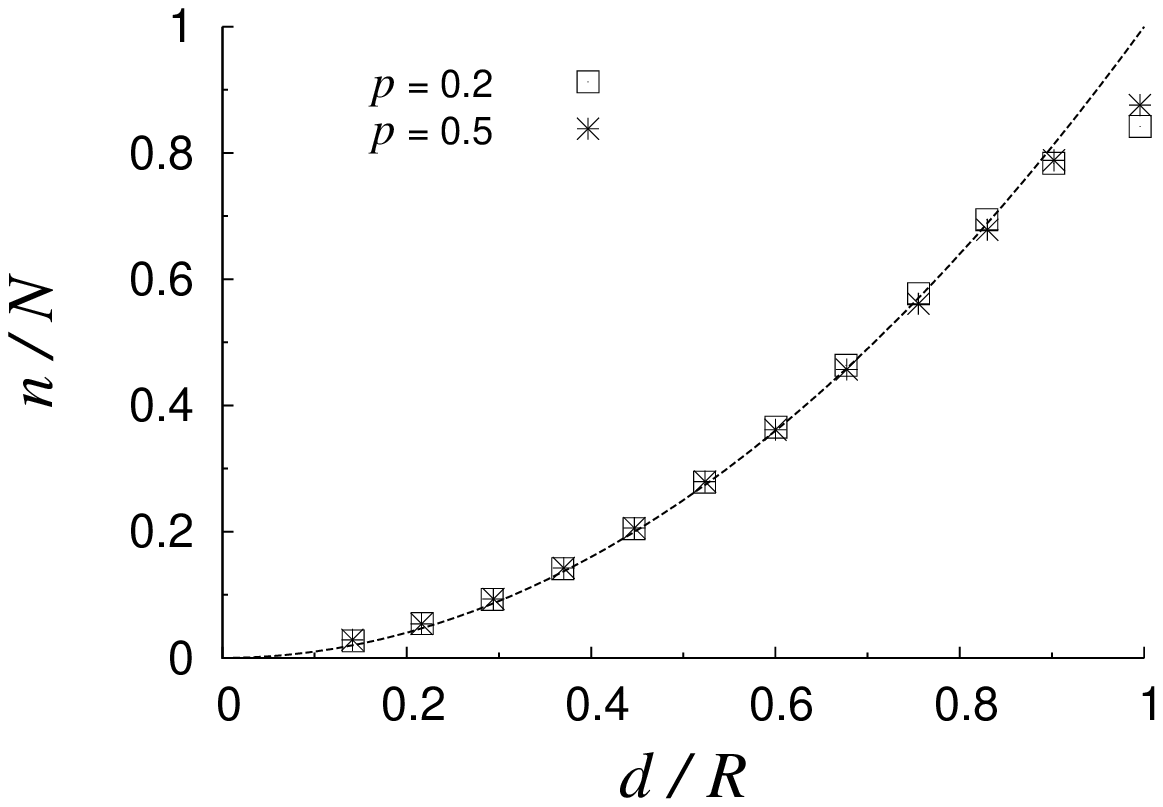} 
\includegraphics[width=\columnwidth,clip]{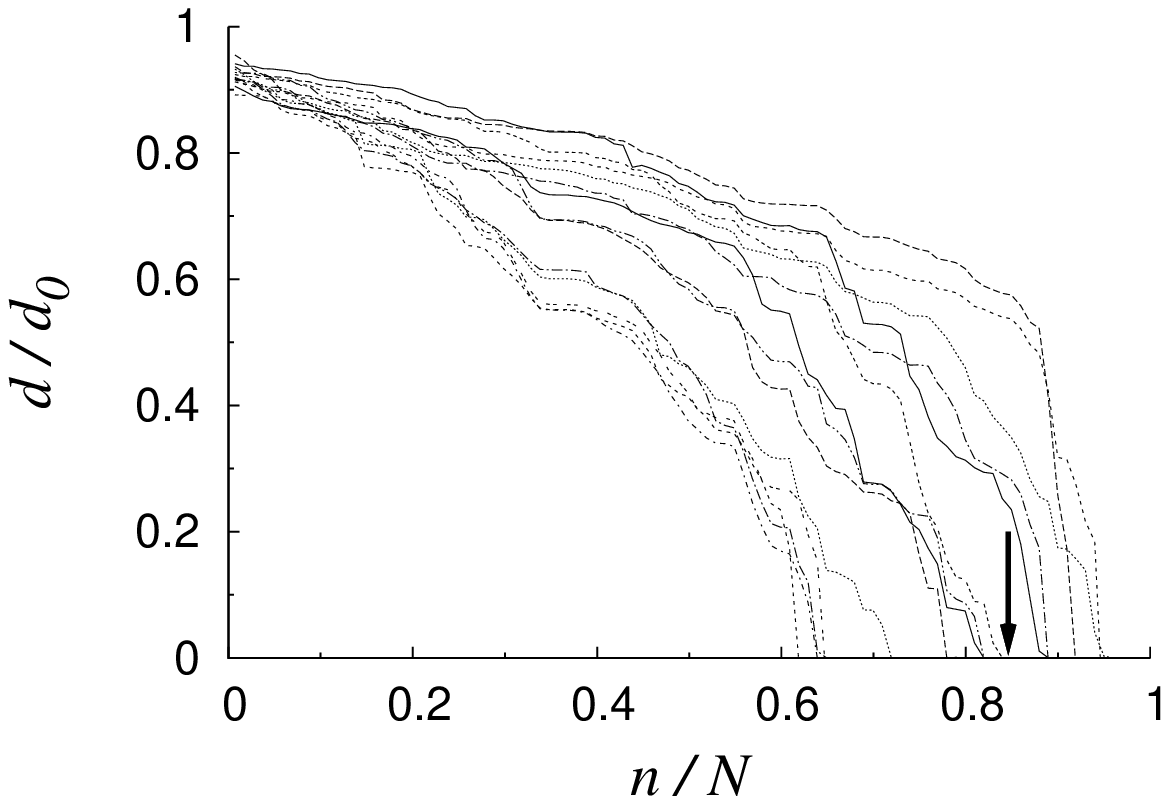} 
\includegraphics[width=\columnwidth,clip]{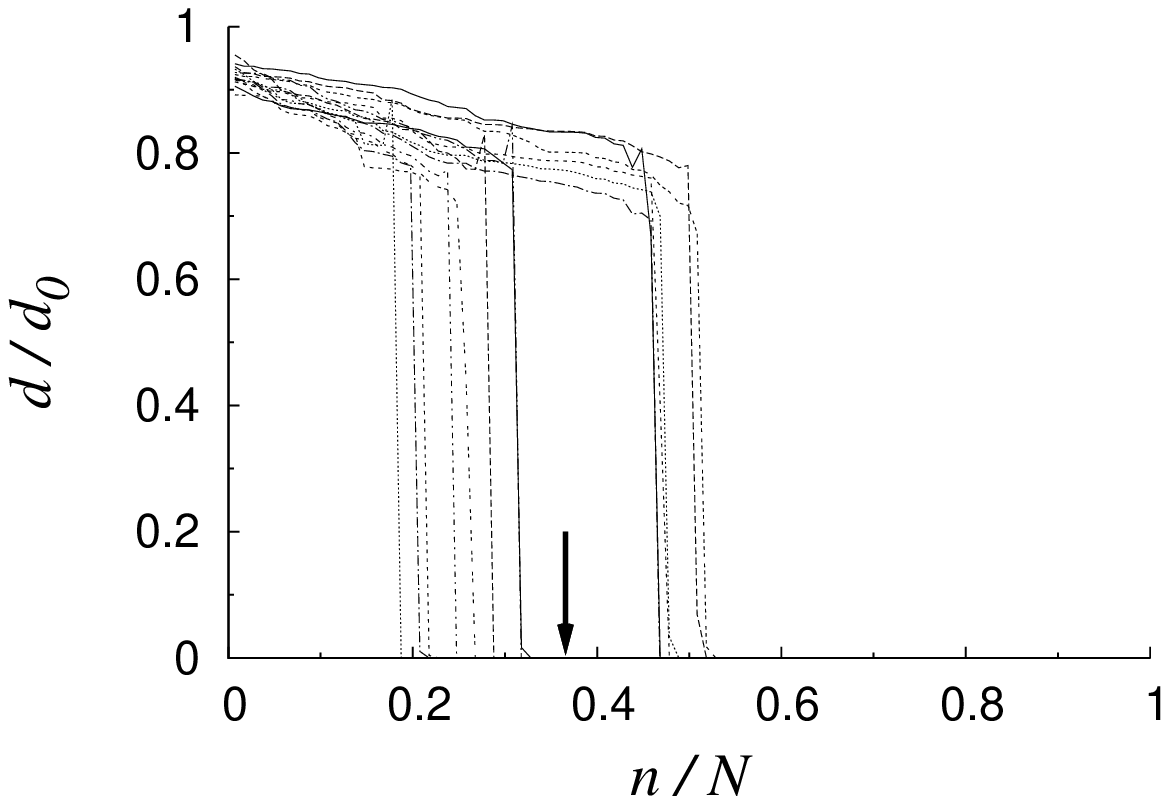} 
\end{center}
\caption{Upper panel: average number of serving steps as a function of initial 
distance from the counter for different values of the probability of 
non-radial moves $p$. The dashed line is a plot of formula~(\ref{e:st2DQ}).
Lower panels: distance from the counter of some individuals lying initially 
in a given shell at distance $d_{0}$ from the origin 
as a function of the serving steps:  $d_{0}=1$ (center) and
$d_{0}=0.61$ (bottom). The arrows indicate the average waiting time from the shells.
Other parameters are: $N=843$, $p = 0.2$, $\phi=0.6$, $\Delta r = 0$ and $K=30$.}
\label{fig.1}
\end{figure}

In the case of a one-dimensional queue of $N$ individuals, it is clear that an
agent initially at a distance $d  \propto n$ from the counter (that is, the $n$-th agent), 
will be served after a time $n$ (in units of serving steps). In two
dimensions, if the same number of people are crowding around the counter,
there will be competition among the agents lying initially within the {\em shell} 
at distance $d$ from the origin.
In this case, the equivalent of the one-dimensional queuing process will be realized if,
from the $d-$shell, each agent will have to wait that all the individuals closer than 
her to the counter be served before she is served. In this case, one should expect 
a number of serving steps given by
\begin{equation}
\label{e:st2DQ}
n_{\rm seq}(d) = N \left( 
           \frac{d}{R}
         \right)^{2}
\end{equation}

\begin{figure*}[t!]
\begin{center}
\includegraphics[width=13 truecm,clip]{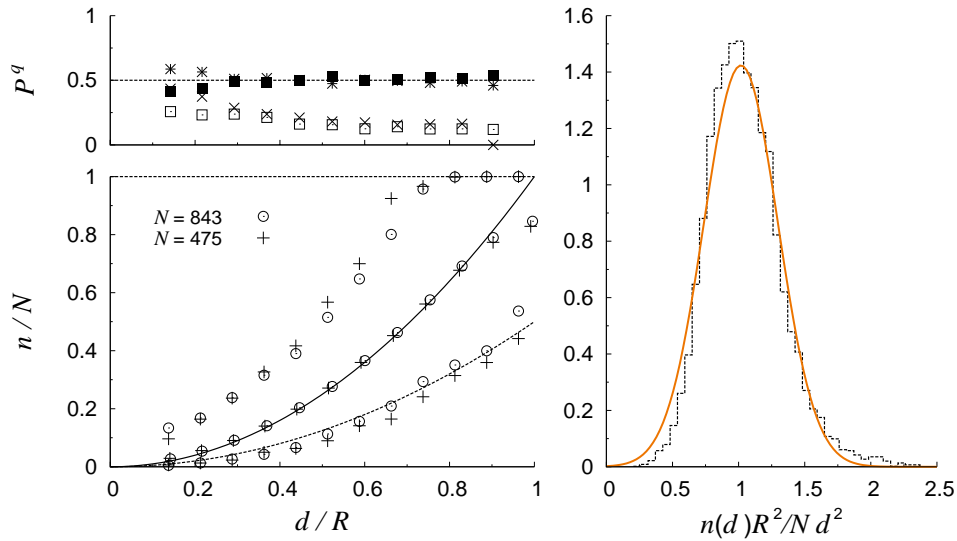} 
\end{center}
\caption{(Color online) Lower left panel: average, maximum and minimum number of 
serving steps as functions of the initial distance from the counter (symbols),
$n_{\rm seq}(d)$ (solid line) and $n_{\rm seq}(d)/2$ (dashed line).
Upper panel:  $\mathcal{P}_{<}^{q=0}$ (filled squares), $\mathcal{P}_{<}^{q=0.25}$ 
(empty squares), $\mathcal{P}_{>}^{q=0}$ (asterisks) and  $\mathcal{P}_{>}^{q=0.25}$ 
(crosses). Right panel: probability density of 
$n(d)/n_{\rm seq}(d)$ (stairs) and normal distribution $N(\mu=1,\sigma=0.28)$ 
(solid line).  Other parameters are: $N=843$, 
$\phi=0.6$, $p = 0.2$, $\Delta r = 0$ and $K=30$.}
\label{fig.2}
\end{figure*}

\subsection{Homogeneous crowds}

The upper plot reported in Fig.~\ref{fig.1} shows that,
under the simple rules of our rearrangement dynamics, $\langle n(d) \rangle = n_{\rm seq}(d)$
for a group of identical individuals. 
This means that, as suggested by simple intuition, on average  there is no difference 
between an ordered one-dimensional queuing process and a chaotic mob formed by an equal 
number of agents crowding around the same serving desk.
Remarkably, this result appears rather insensitive to variations in the 
probability of non-radial moves. This confirms the soundness and robustness of 
our approach. In the case $p=0$ (purely radial moves), we have found that 
the external shells are left virtually motionless, 
as a result of which the crowd is served 
sequentially from the inner shells to the outer ones, more or less 
following the ordered list of distances from the origin in the initial configuration. 
Consequently, the sequential serving law~(\ref{e:st2DQ})  is (trivially) recovered in 
the case $p=0$ too. Indeed, the requirement $p \neq 0$  reflects a realistic 
characteristic of the dynamics  of waiting groups, where individuals share an obvious
bias towards radial moves that makes them approach the counter, but from time to 
time deviate sideways as local bubbles of low density appear under way. 
\\
\indent The data reported in the upper panel of Fig.~\ref{fig.1}
represent averages over many individuals lying initially at the same distance from the counter.
As a matter of fact, in a  two-dimensional arrangement 
different agents from the same shell show amply different 
waiting times, as testified by the two cases reported in the two other panels of the same figure.
It is apparent  that serving times up to about 30 \% below the average 
characterize certain individuals from the same shell, while  at the same 
time others will have to wait longer to be served.
In order to quantify the effect of such fluctuations, in Fig.~\ref{fig.2} 
we plot in one representative case the average, maximum and 
minimum serving times  as functions of the shell distances from the counter. 
We see that the lowest serving times $n_{\rm min}(d)$ 
are well approximated by the same  sequential law~(\ref{e:st2DQ}) 
but corresponding to half the number of agents. 
Hence, the luckiest individuals are served as if they were crowd-queuing
with half the number of people. However, the less fortunate agents wait as if 
they shared  the place with about twice as much agents, with the obvious constraint  that
the maximum serving time cannot exceed $N$. 
It is interesting to remark that the {\em last-of-all} condition ($n_{\rm max}(d) = N$) 
persists for agents initially sitting as close to the counter as $d=0.8 R$.
\\
\indent Additional, more quantitative information can be obtained by
calculating the probability that an agent be served in a number of steps 
below (or above) a given fraction of the sequential-queuing prediction
from a given distance.
Let $\mathcal{N}_d$ indicate the ensemble 
of waiting times of all agents from the $d$-shell over all the 
independent serving runs.
Then, we can measure the probability that an individual is
served in a number of steps lower (or greater) than a given fraction 
$1\mp q$ ($0<q<1$) of  the theoretical expectation, 
Eq.~(\ref{e:st2DQ}), that is 
\begin{equation}
\label{e:Pless}
\mathcal{P}_{\lessgtr}^{q}(d) = \left\langle\left\langle
                          \Theta \left[ 
                          \pm (1 \mp q) n_{\rm seq}(d) 
                          \mp n(d)\right]
                           \right\rangle\right\rangle_{n(d)\in\mathcal{N}_d} 
\end{equation}
where $\Theta(x)$ is the Heaviside step function and 
the operation $\langle\langle \dots \rangle \rangle$ makes the double 
average over shell members and independent runs explicit. 
\\
\indent The results reported in  Fig.~\ref{fig.2} show that the chances of reaching 
the counter in  a number of steps below (or above) the purely sequential prediction ($q=0$) 
are about 50 \%, and to a large  extent insensitive of the initial distance.
Furthermore, the chances of being served faster than  0.75 (or more slowly 
than 1.25) of such expected waiting time are between 15 and 20 \%. 
These observations make more quantitative the intuitive feeling that competing 
in a crowd to be served  can be rewarding as well as imply longer waiting  times 
than in ordered, one-dimensional queues.
Furthermore, the above results suggest that the reduced serving 
times from all shells, $n(d)/n_{\rm seq}(d)$, are described by the same probability density.
The right panel in Fig.~\ref{fig.2}
shows that such curve is  reasonably well~\footnote{To be more precise, the overall 
data set features a skewness  of 0.67  and a kurtosis excess of 1.14.}
approximated by a normal density of unitary mean and standard 
deviation $\sigma = 0.28$. This provides a handy criterion for quantifying to an 
arbitrary confidence level the variability displayed by the serving times. 
For example,  in about $70 \%$ of the cases, an agent starting at a distance $d$ 
will have to wait between $0.7 \, n_{\rm seq}(d)$ and $1.3 \, n_{\rm seq}(d)$.
\\
\indent It should be remarked that the above results concerning minimum and maximum 
waiting times provide ready information under the hypothesis of equally 
spaced serving times $t_{n+1}-t_n=\tau$. More generally, the same predictions are
expected to hold true in the case of uncorrelated, normally distributed serving times
(likely a reasonable description of most common counter dynamics), provided 
the serving and rearrangement time scales are still uncoupled.
In situations where different instances of the distribution $\mathcal{P}(\tau)$ 
are relevant, the relation between 
serving times and number of serving steps could be nonlinear. 
In such cases, the convolution of  the queuing and the serving  
time statistics should be accounted for explicitly
and our algorithm should be modified accordingly.

\subsection{Heterogeneous crowds\label{sec:db}}

We have seen that, on average, an individual involved in a queuing process within an homogeneous 
crowd gets to the counter following the sequential rule, that is after all agents closer 
than her to the origin are served. 
\begin{figure}[t!]
\begin{center}
\includegraphics[width=\columnwidth,clip]{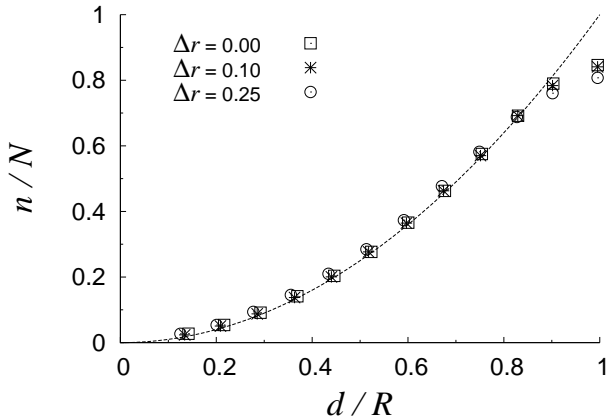} 
\end{center}
\caption{Average number of serving steps as a function of initial distance from the counter
for different values of the size heterogeneity amplitude $\Delta r$.
The dashed line is a plot of formula~(\ref{e:st2DQ}).
Parameters are: $N=843$, $p = 0.2$, $\phi=0.6$, $K=30$.}
\label{fig.new}
\end{figure}
Fig.~\ref{fig.new} proves that such average behavior persists also in the case of a poly-disperse 
group of agents, no matter the spread of the size distribution.  
However, each point in  Fig.~\ref{fig.new}
is worked out  by averaging over a population of agents that share the same initial distance from
the counter but are  heterogenous in size. What happens if all those serving times are 
sorted as functions of the agents' radii?

\begin{figure}[t!]
\begin{center}
\includegraphics[width=\columnwidth,clip]{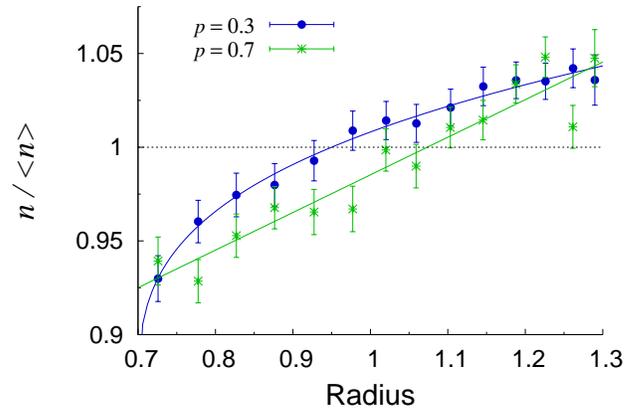} 
\end{center}
\caption{(Color online).  Serving time as a function of the agent size for individuals lying initially 
at a distance $d > 0.9 R$. The serving times have been binned along the radius axis and 
normalized to the average serving time of the considered shell. 
Solid lines are guides to the eye. 
Parameters are $N=475$, $p = 0.3$, $\Delta r = 0.3$ and $K=30$.}
\label{fig.tauvsr}
\end{figure}

Fig.~\ref{fig.tauvsr} provides a clear-cut
answer in the case of individuals initially lying within the outermost 
shell: starting from the same distance, 
the smallest agents get to the counter about 10 \% faster than the bulkiest ones. This phenomenon 
seems to depend to a certain extent on the details of the rearrangement dynamics, although 
the overall advantage displayed by the tiniest agents over larger ones seems not to be
altered by the actual value of the probability of non-radial moves $p$.
We remark that, as expected, the average serving time that characterizes the 
shell (indicated as $\langle n \rangle$ in the figure) also marks the dynamics of 
individuals of average size, that is those with  $r=1$.

\section{Structural properties of the hard-disk crowd\label{sec:3}}

\begin{figure}[t!]
\begin{center}
\begin{tabular}{ll}
\includegraphics[width=3.9 truecm,clip]{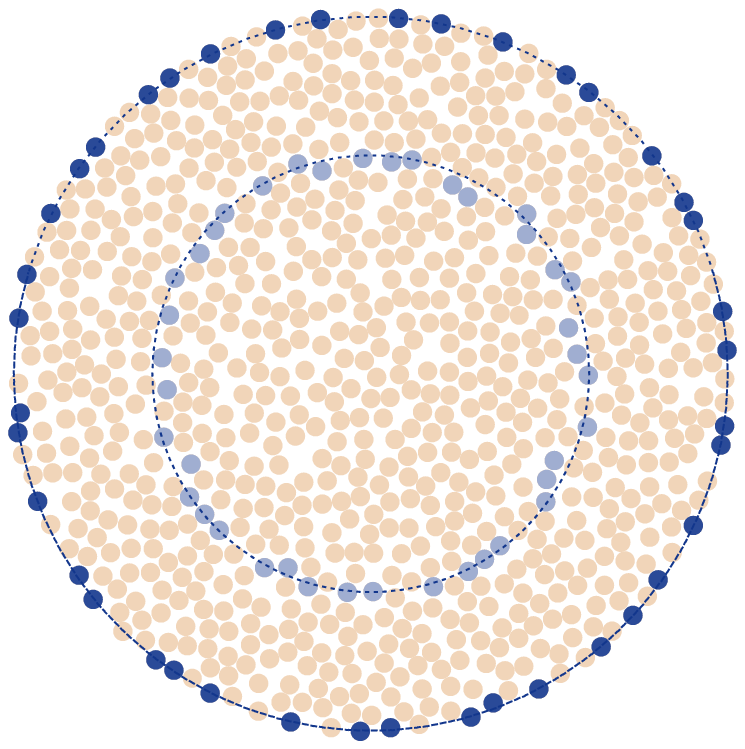} &
\includegraphics[width=3.9 truecm,clip]{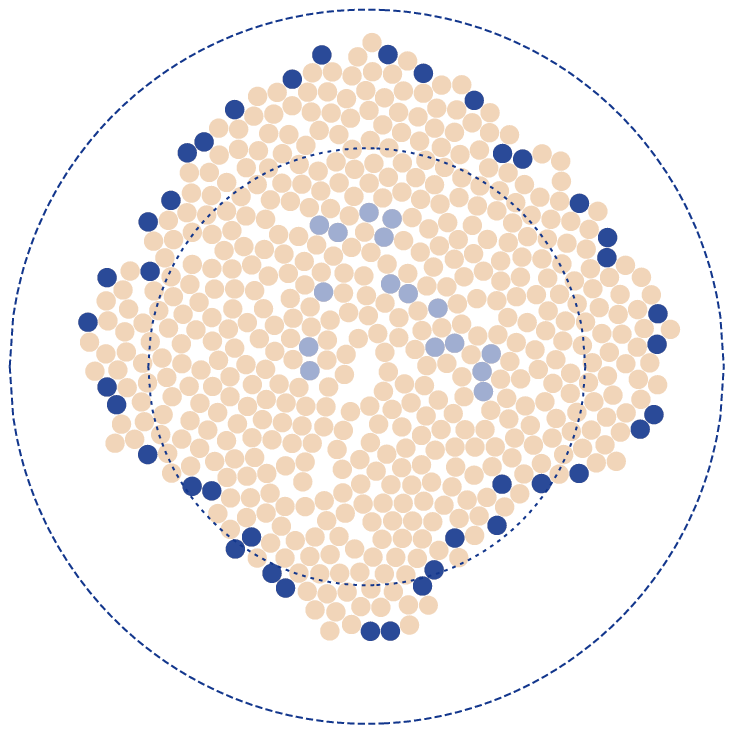} \\
\includegraphics[width=3.9 truecm,clip]{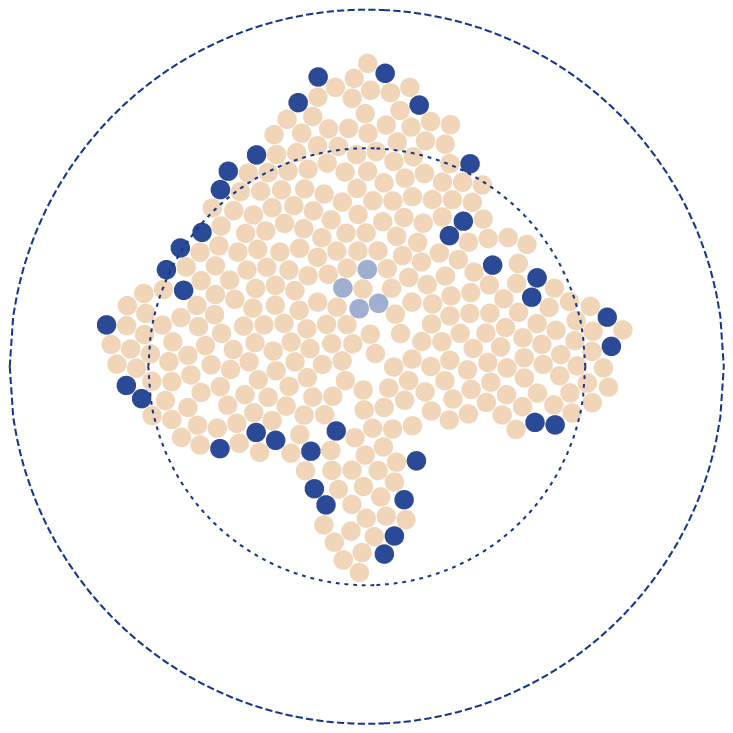} &
\includegraphics[width=3.9 truecm,clip]{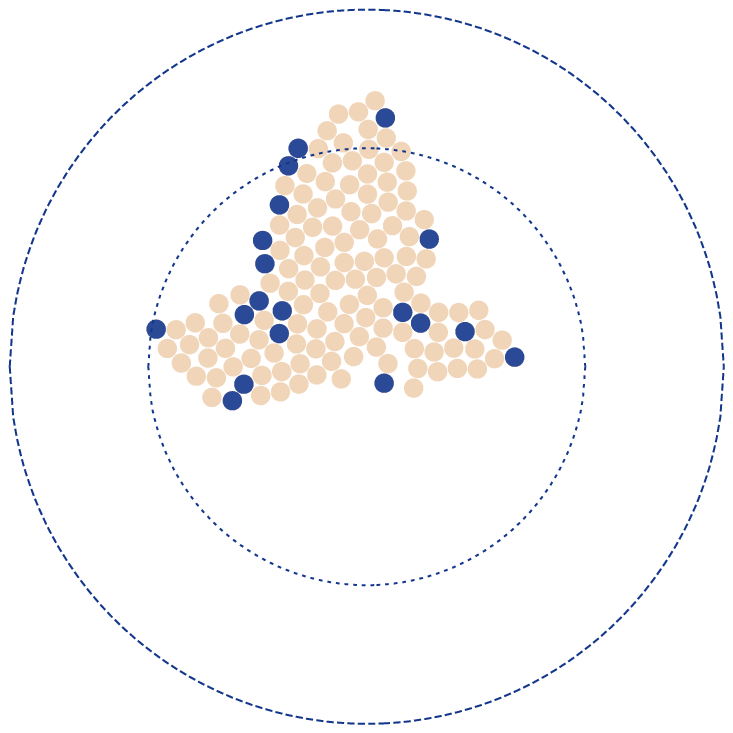} \\
\end{tabular}
\end{center}
\caption{(Color online). Four snapshots during the crowd-queuing process at different steps $n$.
From top to bottom and left to right: initial configuration, $n=300$, 
$n=500$ and $n=700$. The long- and short-dashed lines mark two shells at $d=0.99$ and $d=0.6$, 
while the corresponding agents are represented by dark and light blue circles, respectively. 
Parameters are: $N=843$, $\phi=0.6$, $p=0.2$ and $\Delta r = 0$.}
\label{fig.confs}
\end{figure}

\begin{figure}[t!]
\begin{center}
\includegraphics[width=\columnwidth,clip]{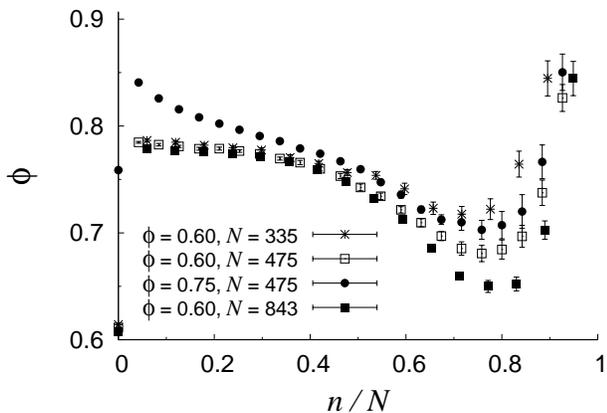} 
\end{center}
\caption{Effective disk area fraction as a function of the serving step during 
the crowd-queuing process. Parameters are $p = 0.2$, $\Delta r = 0$ and $K=30$.}
\label{fig.fract}
\end{figure}

\begin{figure}[t!]
\begin{center}
\includegraphics[width=\columnwidth,clip]{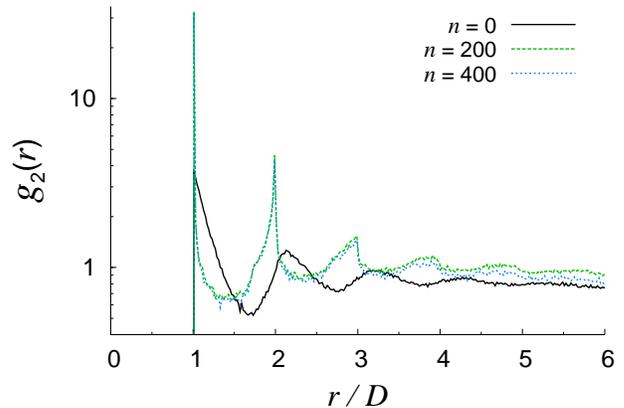} 
\end{center}
\caption{(Color online)  Radial distribution function versus pair-wise distance in units 
of the diameter $D$  at different steps of the serving process 
Other parameters are: $N=843$, $\phi=0.6$, $p=0.2$, $\Delta r = 0$ and $K=30$.}
\label{fig.g2r}
\end{figure}

It is interesting to analyze quantitatively the spatial configurations 
describing our granular crowd systems during their temporal evolution. Fig.~\ref{fig.confs} shows 
four snapshots of the group at different stages for an initial disk area fraction 
$\phi = 0.6$.  Overall, a tendency of the group of agents to deviate 
from the circular shape is observed, likely as a result of non-radial moves 
performed along the way. A similar scenario is confirmed by many other analyzed 
instances (not reported here). Likewise, it appears that the serving process also leads 
to a {\em tightening} of the system.  In order to quantify this
effect, we have calculated the variations of the agents' effective surface coverage  by 
performing Dirichlet tesselation of the configurations at the end of 
each rearrangement stage. Fig.~\ref{fig.fract} reveals that the first phase 
of the process indeed marks a rapid increase of $\phi$ between 10 and 20 \%, depending
on the initial value of the disk area fraction. Subsequently,  
the serving process leads to a loosening of the group, which appears to be more marked the denser the 
starting configuration. Interestingly, towards the end of the process 
the last-served agents seem to organize themselves into tighter structures.
\\
\indent Visual inspection of the systems' patterns suggest that local order appears to
be enhanced late in the serving process (see again Fig.~\ref{fig.confs}). To make this
observation more quantitative, we have calculated the radial distribution function 
$g_2(r)$ (RDF) at different steps in the course of the serving dynamics. The results 
are summarized in Fig.~\ref{fig.g2r} for one typical case. 
The initial configurations, as one should expect, 
reflect the known RDF of hard disk fluids~\cite{Guo:2006xa,Torquato:2002ol}. However, 
as the serving process progresses, short-range translational order  gets manifestly enhanced
alongside with the rapid increase of surface coverage. 
\\
%
\indent It is interesting to observe that the calculated RDFs appear  remarkably similar to 
the ones characterizing the non-equilibrium configurations obtained in Ref.~\cite{Kansal:2000ft}
through a modification of Eden's algorithm targeting a pre-determined degree
of orientational order. In fact, it is instructive to investigate whether the observed enhancement 
of translational order is also accompanied by bond-orientational order. To do this, 
we employed the computed Dirichlet tesselations to evaluate the global 
bond-orientational order parameter

\begin{figure}[b!]
\begin{center}
\includegraphics[width=\columnwidth,clip]{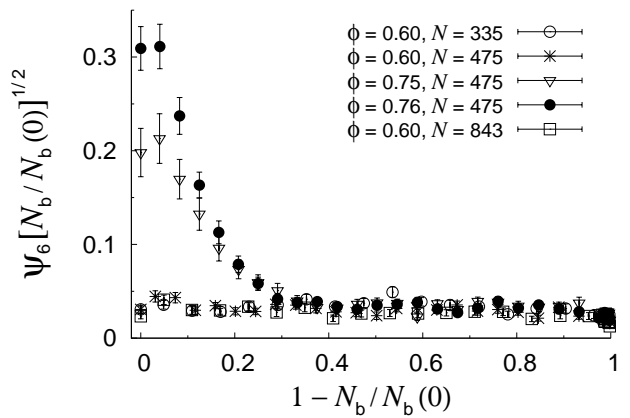} 
\end{center}
\caption{Global bond-orientational order parameter 
versus fraction of removed geometrical bonds during the crowd-queuing process
(proceeding left to right)
for different choices of  $N$ and $\phi$. 
Other parameters are $p = 0.2$, $\Delta r = 0$ and $K=30$.}
\label{fig.psi6}
\end{figure}
\begin{equation}
\label{e:psi6}
\psi_6  = \frac{1}{2N_b} \sum_i \sum_{j=1}^{n_i} \cos (6 \, \theta_{ij})
\end{equation}
where $n_i$ is the number of geometrical neighbors of the $i$-th disk, 
$\theta_{ij}$ is the angle formed by the bond between the $i$-th and $j$-th disks 
in the dual Delaunay triangulation and some 
arbitrary but fixed direction (in our case the $x$ axis) and 
$N_b$ is the total number of such geometrical bonds. By definition, $\psi_6=1$
in the most ordered arrangement, that is the triangular lattice, whereas 
$\psi \propto N_b^{-1/2}$ in a disordered, gas-like 
phase~\footnote{There are exceptions,
essentially due to the {\em global} nature of the definition~(\ref{e:psi6}). However, these 
only concern pathological configurations that are extremely unlikely to occur in
our case~\cite{Kansal:2000ft}.}.
Fig.~\ref{fig.psi6} shows that queuing processes that start at moderate surface coverage 
do not develop appreciable bond-ordering, as $\psi \propto N_b^{-1/2}$ for the whole 
duration of the process. When the initial area fraction is higher, the first configurations indeed 
feature non-negligible bond-orientational order. However, the serving dynamics
invariably brings the system back to gas-phase configurations on a rather short time scale -
when about 20 \% of the agents have been served, any traces of 
bond-orientational order have been already wiped out. 
\\
\indent Overall, the above analysis reveals that the configurations obtained through our 
simple method are typical non-equilibrium 
configurations of hard disks, featuring rather high densities and maximum degree of 
disorder at the same time. Therefore, our method might as well prove 
interesting for non-equilibrium hard-disk packing issues~\cite{Kansal:2000ft}.

%
\section{Conclusions}
%

In this paper we have introduced a simple agent-based Monte Carlo model of crowd dynamics,
with the aim of putting forward a viable and handy alternative to more 
detailed and involved modeling strategies for a variety of crowd dynamics issues. 
Agents are modeled as hard disks and the evolution of the crowd spatial configuration
is reconstructed through a series of two-step moves: in the first step a single 
agent is selected and displaced according to a certain rule, possibly to infinity, 
which amounts to removing her from the game (R1). In the second step, the rest of the 
group is let rearrange through a sequence of Monte Carlo moves obeying a certain 
displacement rule (R2)  until the convergence of a suitable global observable has been achieved.  
The two aforementioned rules R1 and R2 are meant to identify the nature of
the specific crowd dynamics that one wishes to simulate. 

We have shown that out theoretical framework provides results in agreement with 
physical intuition in the problem of {\em crowd-queuing}, namely that of 
a group of agents who gather in a disordered manner around a counter (a ticket-desk, a bar, etc.)
waiting to be served one by one and leaving the place once served.  
Our model shows that on average the crowd reproduces one-dimensional, sequential queuing
for a group of agents of identical size, in agreement with intuition. 
This means that an agent starting from a certain distance $d$
from the counter will have to wait that all agents initially enclosed within
the circular $d$-shell will have been served. 
Obviously, however, a two-dimensional crowd provides 
a queuing environment that can yield slower-than as well as faster-than-sequential serving times 
for a given individual. Our results indicate that the two instances occur with a probability of about 
$1/2$, independently of the initial agents' position.
Remarkably, the above results are rather insensitive to changes in the probability 
of non-radial moves, which testifies to the robustness of our approach. 
\\
%
%
\indent Remarkably, size of agents matters in crowd dynamics processes. 
On average the serving times of an heterogeneous group still follow the intuitive sequential prediction, 
independently of the strength of the agents' size heterogeneity.
However, starting from identical initial distances,
the smaller the agent the shorter her waiting time. The tiniest individuals manage 
to sneak through their fellow queuers more 
effectively, thus reaching the counter the first. It is intriguing to draw
an analogy between our findings, {\em i.e.}  larger disks lagging behind smaller ones, 
and the well-known  {\em Brazil-nut} effect, whereby
the largest particles migrate towards the surface when a granular heterogeneous mixture
is shaken~\cite{Mobius:2001fk}. In fact,  particle size segregation effects have 
already been reported in two-dimensional hard-disk packings~\cite{Jullien:1993uq}.
In the case of shaken mixtures, the bias provided by gravity plays an important role in 
determining the segregation phenomenon. Within the framework of our rearrangement dynamics, 
the agents' {\em mixture} is {\em Monte-Carlo shaken} under the bias of 
preferred origin-pointing displacements. 
In this sense, explanations of the Brazil-nut effect
that focus on infiltration of small particles into voids created underneath larger ones 
during shaking might provide relevant clues to our process. Conversely, the analogy with 
crowd dynamics might prove useful in the rationalization of size effects 
in poly-disperse mixtures. For example, this could be the case of the emergence of indirect attractive 
forces favoring flocculation of larger particles resulting from the interference 
between topological perturbations (depletion wakes) induced by the large 
particles~\cite{Duran:1998kx}.
\\
\indent Finally, we have performed a structural analysis of crowding ensembles. 
We have shown that the rearrangement dynamics results in an  overall 
increase of the agent area fraction,  enhancing at the same time short-range translational 
order in the system as the serving process progresses. On the contrary, orientational order 
is effectively suppressed, even in the case of  high initial packing, leading to rather 
dense and at the same gas-like, highly disordered non-equilibrium configurations, 
reminiscent of those accessible through certain seed-based growth protocols~\cite{Kansal:2000ft}.

%
%
%

%
\end{document}